\documentclass[fleqn,10pt]{wlscirep}

\usepackage{url}
\usepackage{color}

\title{Mapping road network communities for guiding disease surveillance and control strategies}

\author[1,2,*]{Emanuele Strano}
\author[2]{Matheus P. Viana}
\author[4,5]{Alessandro Sorichetta}
\author[4,5,*]{Andrew J. Tatem}
\affil[1]{Department of Civil and Environmental Engineering, Massachusetts Institute of Technology (MIT), Cambridge, MA 02139, USA}
\affil[2]{German Aerospace Center (DLR), German Remote Sensing Data Center (DFD), Oberpfaffenhofen, D-82234 Wessling, Germany}
\affil[3]{IBM Research Brazil, SP, Brazil}
\affil[4]{WorldPop, Department of Geography and Environment, University of Southampton, Highfield, Southampton, UK}
\affil[5]{Flowminder Foundation, Stockholm, Sweden}

\affil[*]{estrano@mit.edu, A.J.Tatem@soton.ac.uk }

\begin{abstract} 
Human mobility is increasing in its volume, speed and reach, leading to the movement and introduction of pathogens through infected travelers. An understanding of how areas are connected, the strength of these connections and how this translates into disease spread is valuable for planning surveillance and designing control and elimination strategies. While analyses have been undertaken to identify and map connectivity in global air, shipping and migration networks, such analyses have yet to be undertaken on the road networks that carry the vast majority of travellers in low and middle income settings. Here we present methods for identifying road connectivity communities, as well as mapping bridge areas between communities and key linkage routes. We apply these to Africa, and show how many highly-connected communities straddle national borders and when integrating malaria prevalence and population data as an example, the communities change, highlighting regions most strongly connected to areas of high burden. The approaches and results presented provide a flexible tool for supporting the design of disease surveillance and control strategies through mapping areas of high connectivity that form coherent units of intervention and key link routes between communities for targeting surveillance.

\end{abstract}

\begin{document}

\flushbottom
\maketitle\thispagestyle{empty}

\section*{Introduction}

The world is continuing to become more connected. The speed and reach of global transport infrastructure is increasing, as well as the numbers of travelers using them. An unintended consequence of this is the increasing transport of pathogens, with more outbreaks becoming global pandemics than ever before \cite{tatem2006global}. Recent pandemics, such as SARS \cite{peiris2004severe}, H5N1 \cite{webster2006h5n1} and H1N1 \cite{trifonov2009geographic} spread rapidly between and within countries through the movement of infected travelers by air, land and sea. Moreover, rising global connectivity is facilitating the increasingly rapid spread of drug resistance\cite{MARAIS20163,lynch2011transit,hupalo2016population}. The growth of transport networks is a key factor in driving the speed and extent of disease spread. While air and shipping networks provide long distance connections, enabling rapid pathogen, host and vector movements \cite{tatem2012air, tatem2014mapping, lemey2014unifying}, the vast majority of shorter distance movements take place over land. This is particularly true in low income settings, where poorer populations are disproportionately affected by infectious diseases and air travel often remains the preserve of wealthier people. Evidence of the importance of regional connectivity through road networks on infectious disease spread is growing \cite{Moustafa2017virome,tatem2012spatialHIV,Faria2014earlyHIV,kraemer2017spread}. The 2015 West Africa Ebola outbreak illustrated how the emergence of a disease in a highly inter-connected region with poor surveillance capacity facilitated rapid spread, compared to previous outbreaks in poorly connected areas \cite{dudas2017virus,wesolowski2014commentary}.
The density, pattern and amount of road in a region is indicative of how populations are physically connected. Dense road networks develop to serve the needs of highly populated regions covering many settlements, facilitating and promoting extensive and regular travel between them \cite{scott2009world,Linard2012Africa}.
In contrast, areas with relatively fewer roads are indicative of lower population densities, lower rates of travel and poorer connectivity. 
Being able to quantify and map these differing types of regions provides potentially valuable information for tackling infectious diseases. 

The study of road network structure has a long tradition. The general approach, which matured as an integration of transportation geography with graph theory \cite{Garrison1962,Haggett1969}, is to model a road system as a network (or graph) and then to study its topological structure. More recently, network science coupled with the availability of large transport datasets, has boosted the study of road networks \cite{Barthelemy2011}. There is now a considerable body of knowledge on road network structure \cite{Strano2012,strano_EPB_2013,Strano170590}, evolution and environmental and societal impact \cite{porta}. 
However, in contrast to other societal networks, the regular and planar nature of road networks precludes the formation of clear communities, i.e. roads that cluster together shaping areas that are more connected within their boundaries than with external roads. 

Highly connected regional communities can promote rapid disease spread within them, but can be afforded protection from recolonization by surrounding regions of reduced connectivity, making them potentially useful intervention or surveillance units \cite{tatem2014integrating,tatem2010international,lynch2011transit}. For isolated areas, a focused control or elimination program is likely to stand a better chance of success than those highly connected to high-transmission or outbreak regions. For example, reaching a required childhood vaccination coverage target in one district is substantially more likely to result in disease control and elimination success if that district is not strongly connected to neighbouring districts where the target has not been met. The identification of ‘bridge’ routes between highly connected regions could also be of value in targeting limited resources for surveillance\cite{wangdi2015chapter}. Moreover, progressive elimination of malaria from a region needs to ensure that parasites are not reintroduced into areas that have been successfully cleared, necessitating a planned strategy for phasing that should be informed by connectivity and mobility patterns \cite{tatem2014integrating}.
Here we develop methods for identifying and mapping road connectivity communities in a flexible, hierarchical way. Moreover, we map ‘bridge’ areas of low connectivity between communities and apply these new methods to the African continent. Finally, we show how these can be weighted by data on disease prevalence to better understand pathogen connectivity, using \emph{P.falciparum} malaria as an example.

%%%%%%%%%%%%%%%%%%%%%%%%%%%%%%%%%%%%%%%%%%%%%%%%%%%%%%%%%%%%%%%%%%%%

\section*{Data}

\subsubsection*{African road network data}

Data on the African road network (ARN) were obtained from GPS navigation and cartography as described in a previous study \cite{Strano170590}. The dataset maps primary and secondary roads across the continent, and while it does have commercial restrictions, it is a more complete and consistent dataset than alternative open road datasets (e.g. OpenStreetMap \cite{OSM}, gRoads \cite{GSM}. Visual inspection and comparison between the ARN and other spatial road inventories validated the improved accuracy and consistency of ARN, however a quantitative validation analysis was not possible due to the lack of consistent ground-truth data at continental scales. Figure 1a shows the African road network data used in this analysis. The road network is a commercial restricted product and requests for it can be directly addressed to GARMIN \cite{GARMIN}.

\subsubsection*{Plasmodium falciparum malaria prevalence and population maps}

To demonstrate how geographically referenced data on disease occurrence or prevalence can be integrated into the approaches outlined, gridded data on \emph{Plasmodium falciparum} malaria prevalence were obtained from the Malaria Atlas Project (\url{http://www.map.ox.ac.uk/}). These represent modelled estimates of the prevalence of \emph{P.falciparum} parasites in 2015 per 5x5km grid square across Africa \cite{bhatt2015effect}. Additionally, gridded data on estimated population totals per 1x1km grid square across Africa in 2015 were obtained from the WorldPop program (\url{http://www.worldpop.org/}). The population data were aggregated to the same 5x5km gridding as the malaria data, and then multiplied together to obtain estimates of total numbers of \emph{P.falciparum} infections per 5x5km grid square.

%%%%%%%%%%%%%%%%%%%%%%%%%%%%%%%%%%%%%%%%%%%%%%%%%%%%%%%%%%%%%%%%%%%%

\section*{Results}

\subsection*{Detecting communities in the African road network}

We modeled the ARN as a 'primal' road network, where roads are links and road junctions are nodes \cite{Porta2006}. Spatial road networks have, as any network embedded in two dimensions, physical spatial constraints that impose on them a grid-like structure. In fact, the ARN primal network is composed of $300,306$ road segments that account for a total length of $2,304,700 km$, with an average road length of $7.6km \pm 13.2km$. Such large standard deviations, as already observed elsewhere \cite{masucci2009random,strano2013urban,Strano170590}, are due to the long tailed distribution of road lengths, as illustrated in Figure 1c.
Another property of road network structure is the frequency distribution of the degree of nodes, defined as the number of links connected to each node. Most networks in nature and society have a long tail distribution of node degree, implying the existence of hubs (nodes that connect to a large amount of other nodes) \cite{Barthelemy2011}, with the majority of nodes connecting to very few others. For road networks, however, the degree distribution strongly peaks around 3, indicating that most of the roads are connected with two other roads. The long tail distribution of the length of road segments, coupled with the peaked degree distribution, indicates the presence of translational invariant grid-like structure, in which road density smoothly varies among regions while their connectivity and structure does not. Within such grid-like structures it is very difficult to identify clustered communities, i.e. groups of roads that are more connected within themselves than to other groups. This observation is confirmed by the spatial distribution of betweenness centrality (\emph{Bc}), which measures the amount of time the shortest paths between each couple of nodes pass through a road. The probability distribution of \emph{Bc} is long tailed (Figure 1d), while its spatial distribution spreads across the entire network, with a structural backbone form, as shown in Figure 1b. Again, under such conditions and because of the absence of bottlenecks, any strategy to detect communities that employs pruning on \emph{Bc} values \cite{clauset2004finding}, will be minimally effective.

\begin{figure}[!ht]
\centering\includegraphics[width=\linewidth]{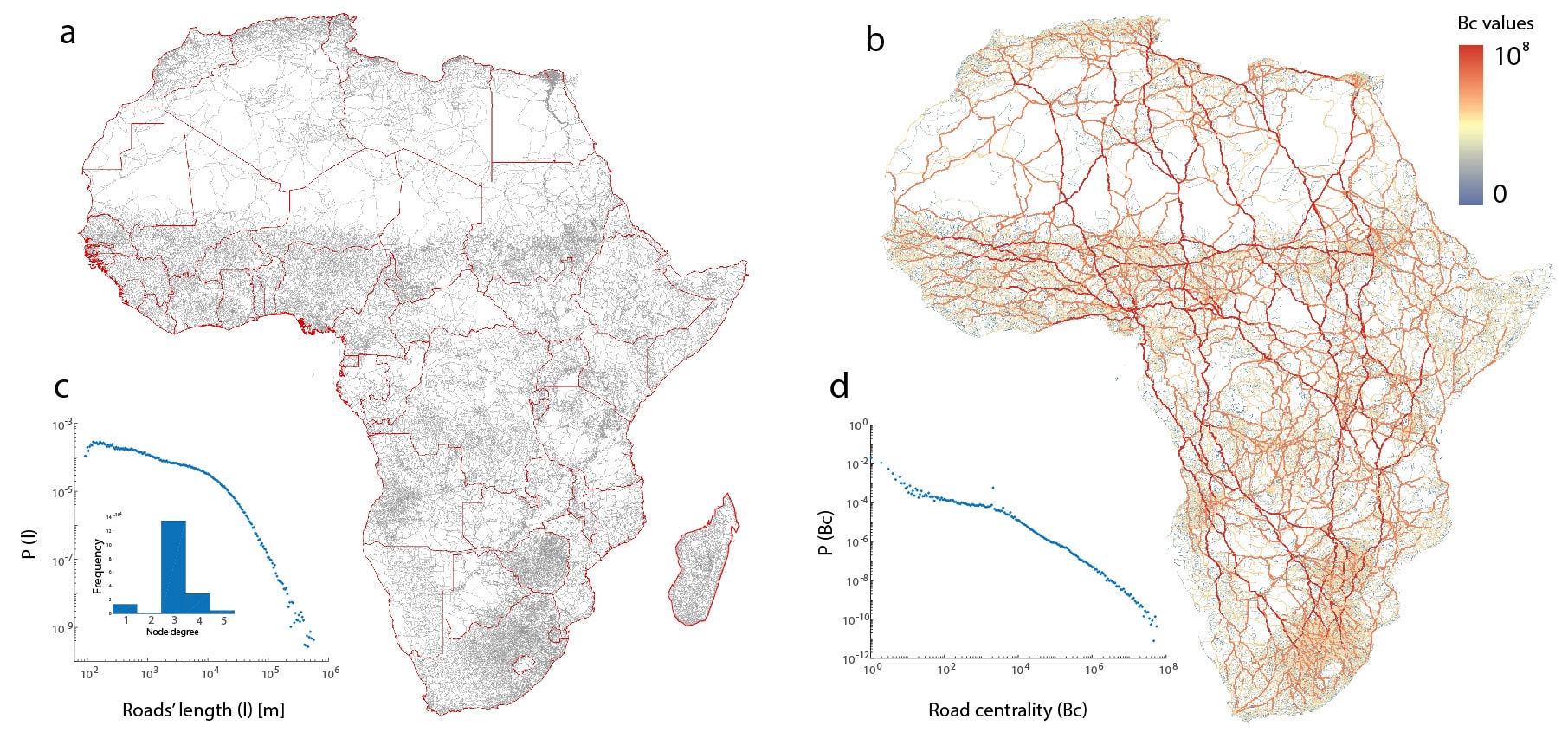}
\caption{ Road network data and features. (a) Africa road network dataset, with inset graph showing node degree frequency and road length distribution plots. (b) Betweenness of each road segment in Africa, with distribution plot in inset graph }
\label{Fig:Data}
\end{figure}

To detect communities in road networks we follow the observation that human displacement in urban networks is guided by straight lines \cite{Rosvall2005}. Therefore, geometry can be used to detect communities of roads by assuming that people tend to move more along streets than between between streets. We developed a community detection pipeline that converts a primal road network, where roads are links and roads junction are nodes \cite{Porta2006}, to a dual network representation, where link are nodes and street junction link between nodes \cite{Porta2006B}, by mean of straightness and contiguity of roads. It is important to note here that the units of analysis are road segments, which here are typically short and straight between intersections, making the straightness assumption valid. Community detection in the dual network is then performed using a modularity optimization algorithm \cite{Newman06062006}.The communities found in the dual network are then mapped back to the original primal road network. These communities encode information about the geometry of road pattern but can also incorporate weights associated with a particular disease to guide the process of community detection.

Nodes in the dual network represent lines in the primal network. The conversion from primal to dual is done by using a modified version of the algorithm known as \emph{continuity negotiation} \cite{Porta2006B}. In brief, we assume that a pair of adjacent edges belongs to the same street if the angle $\theta$ between these edges is smaller than $\theta_c=30^{\circ}$. We also assume that the angle between two adjacent edges $(i,j)$ and $(j,p)$ is given by the dot product $\cos(\theta)=|\mathbf{r}_{i,j}\cdot \mathbf{r}_{j,p}|/r_{i,j}r_{j,p}$, where $\mathbf{r}_{i,j}=\mathbf{r}_j-\mathbf{r}_i$. Under these assumptions, the angle between two edges belonging to a perfect straight line is zero, while it assumes a value of $90^{\circ}$ for perpendicular edges. Our algorithm starts searching for the edge that generates the longest road in the primal space, as can be seen in Figure 2a. Then, a node is created in the dual space and assigned to this road. Next, we search for the edge that generates the second longest road, and a new node is created in the dual space and assigned to this road. If there is at least one interception between the new road and the previous one, we connect the respective nodes in the dual space. The algorithm continues until all the edges in the primal space are assigned to a node in the dual space, as shown in Figure 2b. Note that the conversion from primal to the dual road network has been used extensively to estimate human perception and movement along road networks (Space syntax, see \cite{Rosvall2005}), which also supports our use of road geometry to detect communities.
 
Despite the regular structure of the network in the primal space, the topology of these networks in the dual space is very rich. For instance the degree distribution in dual space follows the power-law $P(k)\sim k^{-\gamma}$. This property has been previously identified in urban networks \cite{Porta2006} and it is strongly related to the long tailed distribution of road lengths in these networks (see Figure 1c). Since most of the roads are short, most of the nodes in dual space will have a small number of connections. On the other hand, there are a few long roads (Figure 2a) that originate at hubs in the dual space (Figure 2b). Our approach for detecting communities in road networks consists then in performing classical community detection in the dual representation (Figure 2c) and then bringing the result back to the primal representation, as shown in Figure 2d. The algorithm used to detect the communities is the modularity-based algorithm by Clauset and Newman \cite{clauset2004finding}.

\begin{figure}[!ht]
\centering\includegraphics[width=\linewidth]{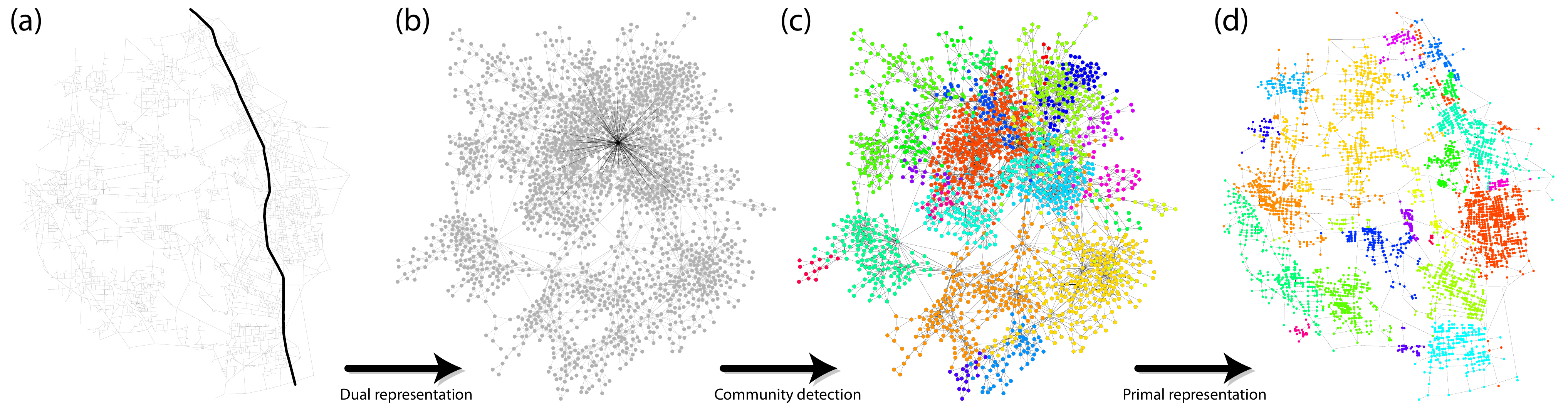}
\caption{ Visualization of the community detection method developed. (a) The primal road network with a single segment highlighted in black. (b) The dual representation of the same network with the single segment highlighted in black. (c) The result of community detection on the dual network representation. (d) Conversion back to the primal road network with the identified communities mapped}
\label{Fig:Method}
\end{figure}

The hierarchical mapping of communities on the African road network, with outputs for 10, 20, 30 and 40 sets of communities, is shown in Figure 3. The maps highlight how connectivity rarely aligns with national borders, with the areas most strongly connected through dense road networks typically straddling two or more countries. The hierarchical nature of the approach is illustrated through the breakdown of the 10 large regions in Figure 3a into further sub-regions in b,c and d, emphasizing the main structural divides within each region in mapped in 3a. Some large regions appear consistently in each map, for example, a single community spans the entire north African coast, extending south into the Sahara. South Africa appears as wholly contained within a single community, while the horn of Africa containing Somalia and much of Ethiopia and Kenya in consistently mapped as one community. The four maps shown are example outputs, but any number of communities can be identified. The clustering that maximises modularity produces 104 communities, and these are mapped in Figure 4. 

Even with division into 104 communities, the north Africa region remains as a single community, strongly separated from sub-Saharan Africa by large bridge regions. South Africa also remains as almost wholly within its own community, with Somalia and Namibia showing similar patterns. The countries with the largest numbers of communities tend to be those with the least dense infrastructure equating to poor connectivity, such as DRC and Angola, though West Africa also shows many distinct clusters, especially within Nigeria. Apart from the Sahara, the largest bridge regions of poor connectivity are located across the central belt of sub-Saharan Africa, where population densities are low and transport infrastructure is both sparse and often poor. The communities mapped in figures 3 and 4 align in many cases with recorded population and pathogen movements. For example, the broad southern and eastern community divides match well those seen in HIV-1 subtype analyses \cite{tatem2012spatialHIV} and community detection analyses based on migration data \cite{tatem2010international}. At more regional scales, there also exist similarities with prior analyses based on human and pathogen movement patterns. For example, the western, coastal and northern communities within Kenya in Figure 4b, identified previously through mobile phone and census derived movement data \cite{wesolowski2013use, wesolowski2012quantifying}. Further, Guinea, Liberia and Sierra Leone typically remain mostly within a single community in Figure 3, with some divides evident in Figure 4c. This shows some strong similarities with the spread of Ebola virus through genome analysis \cite{dudas2017virus}, particularly the multiple links between rural Guinea and Sierra Leone, though Figure 4c highlights a divide between the regions containing Conakry and Freetown when Africa is broken into the 104 communities. Figure 3 highlights the connections between Kinshasa in western DRC and Angola, with the recent yellow fever outbreak spreading within the communities mapped. Figure 4d shows the 'best' communities map for an area of southern Africa, and the strong cross-border links between Swaziland, southern Mozambique and western South Africa are mapped within a single community, as well as wider links highlighted in Figure 3, matching the travel patterns found from Swaziland malaria surveillance data \cite{tejedor2017travel}.

\begin{figure}[!ht]
\centering\includegraphics[width=\linewidth]{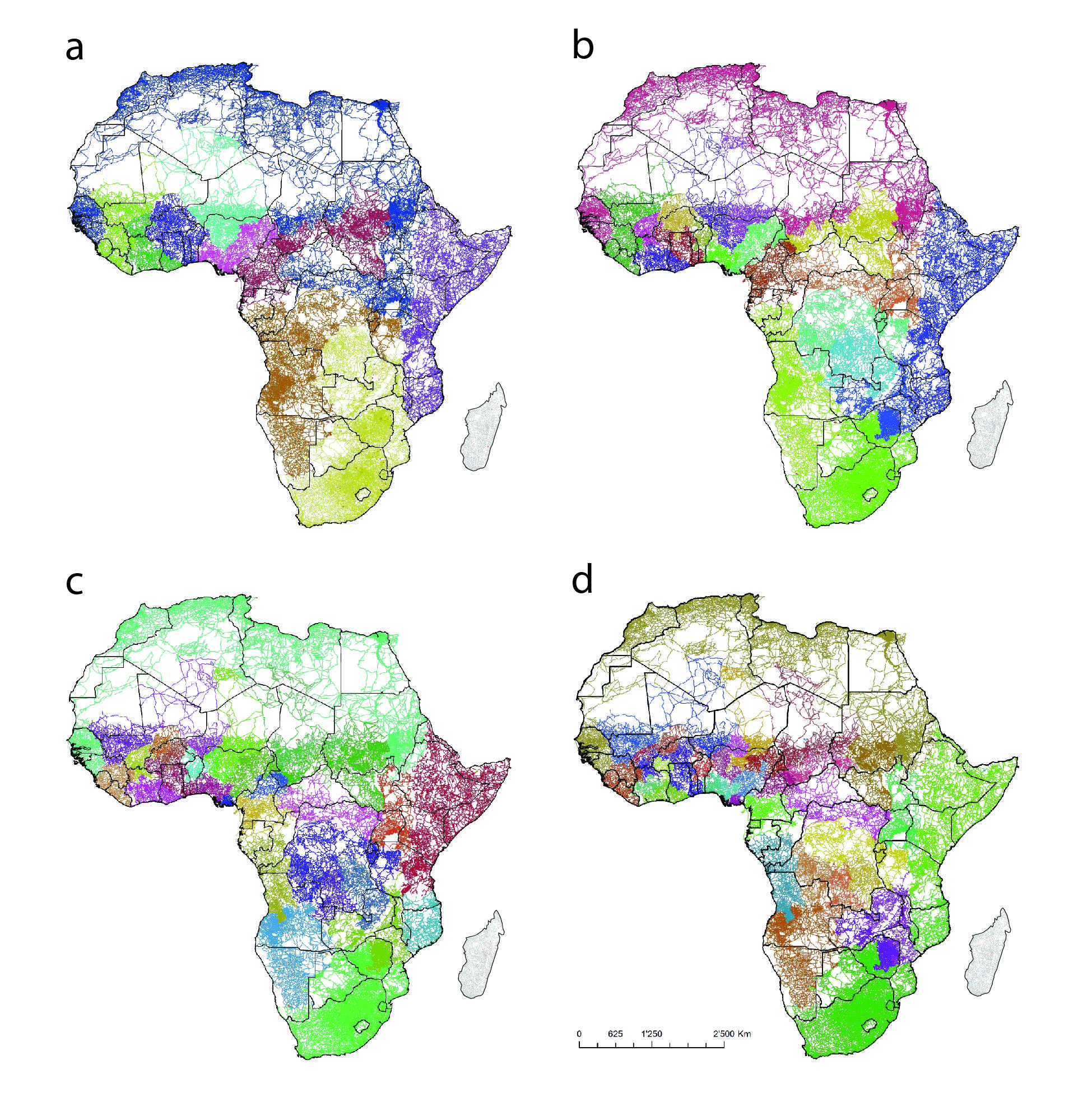}
\caption{ Example outputs of community detection on the unweighted Africa road network, constrained to (a) 10 communities; (b) 20 communities; (c) 30 communities and (d) 40 communities.}
\label{Fig:Clustering}
\end{figure}

\begin{figure}[!ht]
\centering\includegraphics[width=\linewidth]{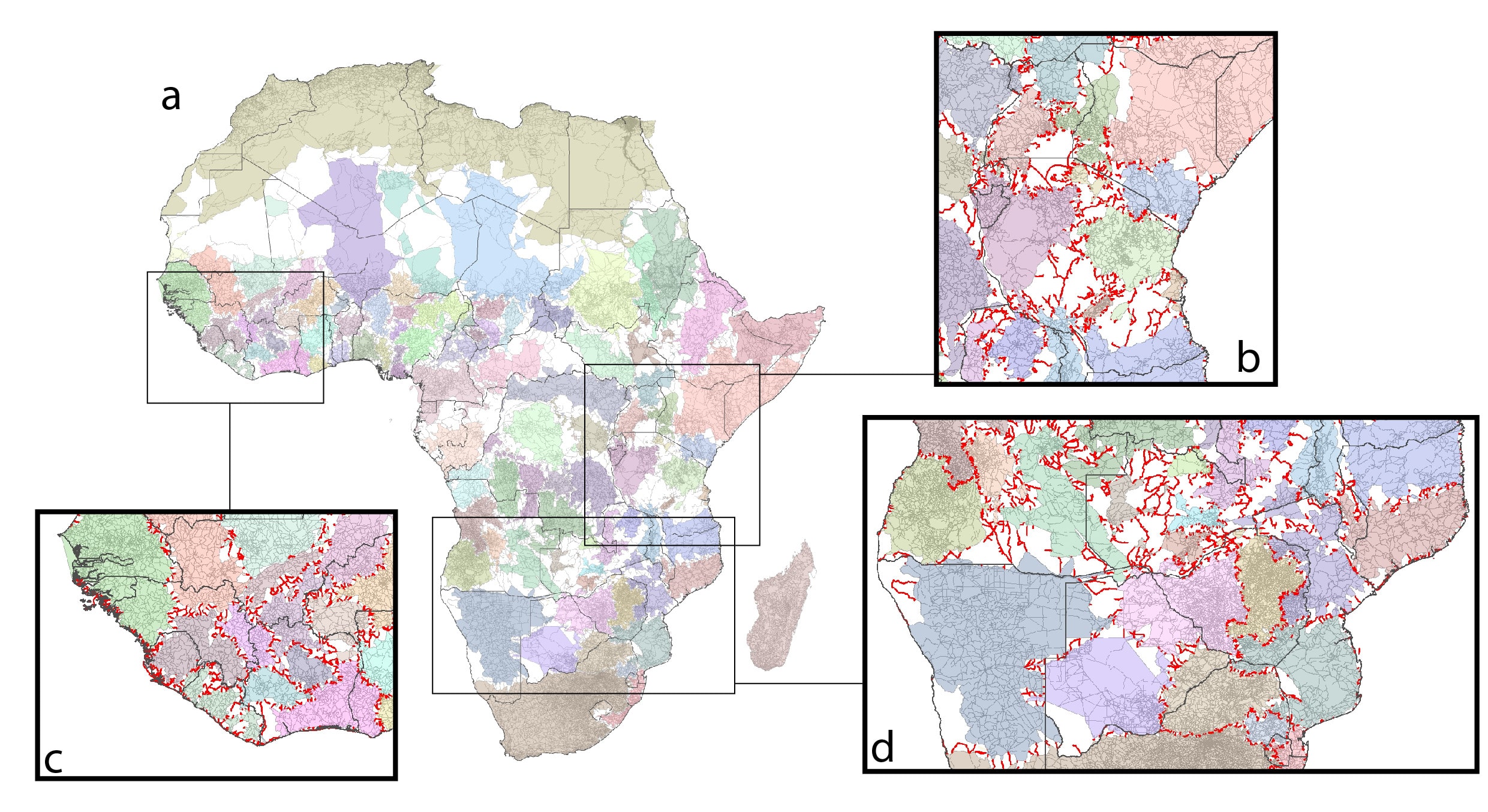}
\caption{Africa road network community mapping for the strongest configuration of communities, with bridge areas of low connectivity identified, for (a) Africa, with bridge areas in white, (b) East Africa with bridge roads between communities mapped in red, (c) West Africa with bridge roads between communities mapped in red, and (d) Southern Africa with bridge roads between communities mapped in red.}
\label{Fig:Bridges}
\end{figure}

%%%%%%%%%%

\subsubsection*{Integrating \emph{P. falciparum} malaria prevalence and population data with road networks for weighted community detection}

The previous section outlined methods for community detection on unweighted road networks. To integrate disease occurrence, prevalence or incidence data for the identification of areas of likely elevated movement of infections or for guiding the identification of operational control units, an adaptation to weighted networks is required. We demonstrate this through the integration of the data on estimated numbers of \emph{P.falciparum} infections per 5x5km grid square into the community detection pipeline. The final pipeline for community detection calculated a trade-off between form and function of roads in order to obtain a network partition.

The form is related to the topology of the road network and is taken into account during the primal-dual conversion. The topological component guarantees that only neighbor and well connected locations could belong to the same community. The functional part, on the other hand, is calculated by the combination of estimated \emph{P. falciparum} malaria prevalence multiplied by population to obtain estimated numbers of infections, as outlined above. 

%The combination of these two factors is important to give the same importance to regions with high malaria prevalence and low population and areas with low prevalence but very populated.

The two factors were combined to form a weight to each edge of our primal network. The weight $w_{i,j}$ of edge $(i,j)$ is defined as

\begin{equation}
	w_{i,j} = \frac{1}{|\mathbf{r}_{i,j}|}\int_{\mathbf{r}_i}^{\mathbf{r}_j}m(\mathbf{r})p(\mathbf{r})d\mathbf{r}
\end{equation}

where $m(\mathbf{r})$ is the \emph{P.falciparum} malaria prevalence and $p(\mathbf{r})$ is the population count, both at coordinate $\mathbf{r}$. These values are obtained directly from the data. When the primal representation is converted into its dual version, the weights of primal edges, given by Eq. 1, are converted into weights of dual nodes, which are defined as

\begin{equation}
	\lambda_{\bar{i}} = \max(w_{i,j}),\quad(i,j)\in\Omega_{\bar{i}},
\end{equation}

where $\bar{i}$ represents the $i-$th dual node and $\Omega_{\bar{i}}$ represents the set of all the primal edges that were combined together to form the dual node $\bar{i}$ (see Fig. 2a-b). Finally, weights for the dual edges are created from the weights of dual nodes, by simply assuming

\begin{equation}
	\lambda_{\bar{i,j}} = \max(\lambda_{\bar{i}},\lambda_{\bar{j}}).
\end{equation}

The dual network weighted by values of $\lambda_{\bar{i,j}}$ was used as input for a weighted community detection algorithm. Ultimately, when the communities detected in the dual space are translated back to primal space, we have that neighbor locations with similar values of estimated \emph{P.falciparum} infections belong to the same communities. For the example of \emph{P. falciparum} malaria used here, the max function was used, representing maximum numbers of infections on each road segment in 2015. This was chosen to identify connectivity to the highest burden areas. Areas with large numbers of infections are often 'sources', with infected populations moving back and forward from them spreading parasites elsewhere \cite{lynch2011transit,pindolia2012human}. Therefore, mapping which regions are most strongly connected to them is of value. Alternative metrics can be used however, depending on the aims of the analyses.

The integration of \emph{P.falciparum} malaria prevalence and population (Figure 5a) through weighting road links by the maximum values across them produces a different pattern of communities (Figure 5b) to those based solely on network structure (Figure 3). The mapping of 20 communities is shown here, as it identifies key regions of known malaria connectivity, as outlined below. The mapping shows areas of key interest in malaria elimination efforts connected across national borders, such as much of Namibia linked to southern Angola \cite{smith2017malaria}, but the Zambezi region of Namibia more strongly linked to the community encompassing neighbouring Zambia, Zimbabwe and Botswana \cite{simon2013malaria}. In Namibia, malaria movement communities identified through the integration of mobile phone-based movement data and case-based risk mapping \cite{tatem2014integrating} show correspondence in mapping a northeast community. Moreover, Swaziland is shown as being central to a community covering, southern Mozambique and the malaria endemic regions of South Africa, matching closely the origin locations of the majority of internationally imported cases to Swaziland and South Africa \cite{raman2016reviewing, tejedor2017travel, koita2013targeting}. The movements of people and malaria between the highlands and southern and western regions of Uganda, and into Rwanda \cite{lynch2015association}, also aligns with the community patterns shown in Figure 5b. Finally, though quantifying different factors, the analyses show a similar east-west split to that found in analyses of malaria drug resistance mutations \cite{pearce2009multiple,lynch2011transit} and malaria movement community mapping \cite{tatem2010international}.

\begin{figure}[!ht]
\centering\includegraphics[width=\linewidth]{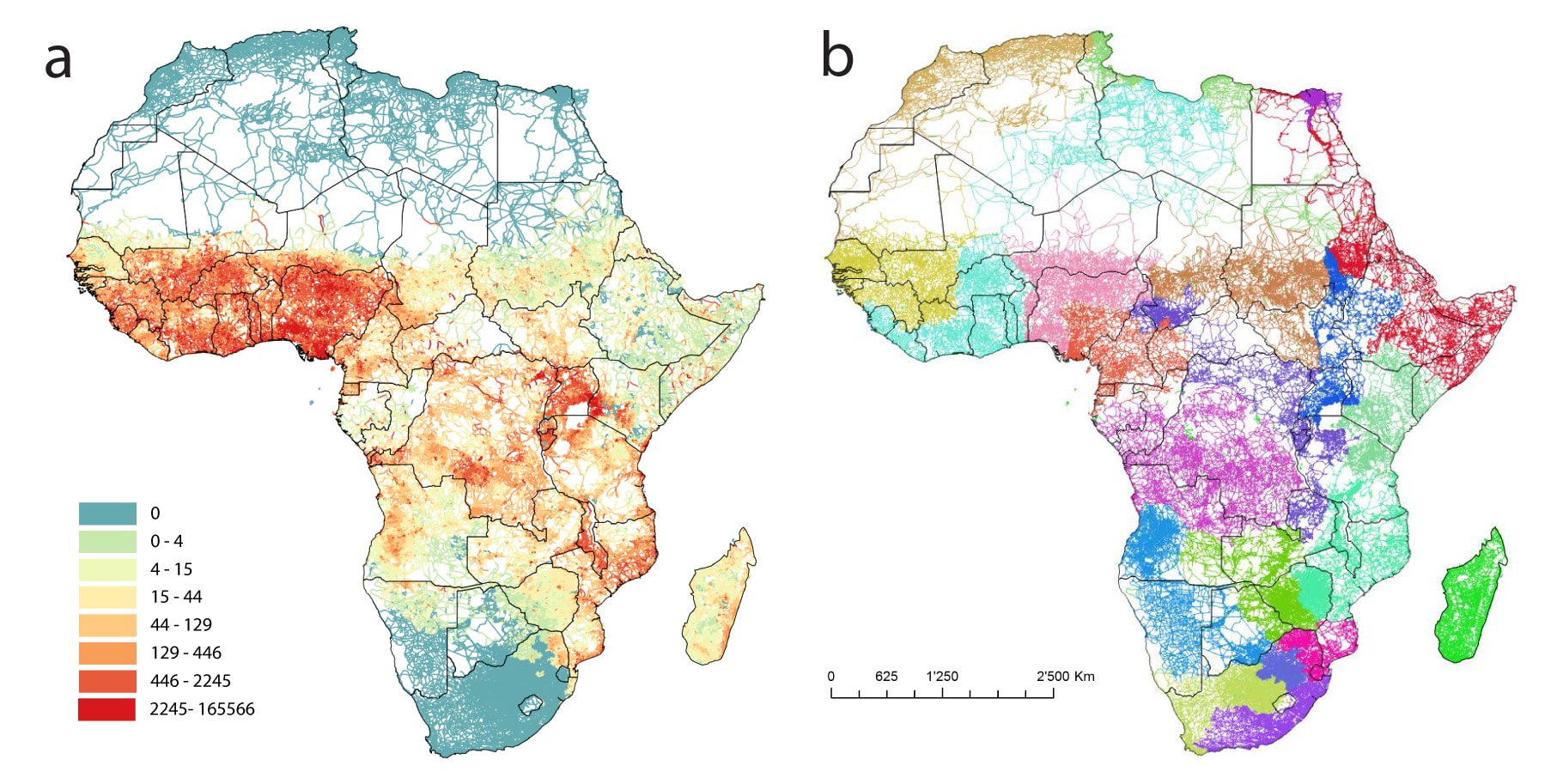}
\caption{ Data and example outputs for \emph{Plasmodium falciparum} malaria weighted road network analyses. (a) Africa road network with each road segment coloured by its maximum value of \emph{P.falciparum} prevalence multiplied by population. (b) Output of community detection on the data in (a), showing the result for 20 communities.}
\label{Fig:Malaria}
\end{figure}

\section*{Discussion}

The emergence of new disease epidemics is becoming a regular occurrence, and drug and insecticide resistance are continuing to spread around the world. As global, regional and local efforts to eliminate a range of infectious diseases continue and are initiated, an improved understanding of how regions are connected through human transport can therefore be valuable. Previous studies have shown how clusters of connectivity exist within the global air transport network \cite{guimera2005worldwide, kaluza2010complex} and shipping traffic network \cite{kaluza2010complex}, but these represent primarily the sources of occasional long-distance disease or vector introductions \cite{tatem2006global,tatem2012air}, rather than the mode of transport that the majority of the population uses regularly. The approaches presented here focused on road networks provide a tool for supporting the design of disease and resistance surveillance and control strategies through mapping (i) areas of high connectivity where pathogen circulation is likely to be high, forming coherent units of intervention; (ii) areas of low connectivity between communities that form likely natural borders of lower pathogen exchange; (iii) key link routes between communities for targetting surveillance efforts.
 
The outputs of the analyses presented here highlight how highly connected areas consistently span national borders. With infectious disease control, surveillance, funding and strategies principally implemented country by country, this emphasises a mismatch in scales and the need for cross-border collaboration. Such collaborations are being increasingly seen, for example with countries focused on malaria elimination (e.g.\cite{APMEN,SADC}), but the outputs here show that the most efficient disease elimination strategies may need to reconsider units of intervention, moving beyond being constrained by national borders. Results from the analysis of pathogen movements elsewhere confirm these international connections (e.g. \cite{lynch2011transit, pearce2009multiple, tejedor2017travel,tatem2012spatialHIV}, building up additional evidence on how pathogen circulation can be substantially more prevalent in some regions than others. 

The approaches developed here provide a complement to other approaches for defining and mapping regional disease connectivity and mobility \cite{tatem2014mapping}. Previously, census-based migration data has been used to map blocks of countries of high and low connectivity \cite{tatem2010international}, but these analyses are restricted to national-scales and cover only longer-term human mobility. Efforts are being made to extend these to subnational scales \cite{sorichetta2016mapping, ruktanonchai2016census}, but they remain limited to large administrative unit scales and the same long timescales. Mobile phone call detail records (CDRs) have also been used to estimate and map pathogen connectivity \cite{wesolowski2012quantifying, tatem2014integrating},
but the nature of the data mean that they do not include cross-border movements, so remain limited to national-level studies. An increasing number of studies are uncovering patterns in human and pathogen movements and connectivity through travel history questionnaires (e.g. \cite{marshall2016key, tejedor2017travel, lynch2015association, bradley2015infection}), resulting in valuable information, but typically limited to small areas and short time periods.
 
There exist a number of limitations to the methods and outputs presented here that future work will aim to address. Firstly, the hierarchies of road types are not currently taken into account in the network analyses, meaning that a major highway and small local roads contribute equally to community detection and epidemic spreading. The lack of reliable data on road typologies, and inconsistencies in classifications between countries, makes this challenging to incorporate however. Moreover, the relative importance of a major road versus secondary, tertiary and tracks is exceptionally difficult to quantify within a country, let alone between countries and across Africa. Finally, data on seasonal variations in road access does not exist consistently across the continent. Our focus has therefore been on connectivity, in terms of how well regions are connected based on existing road networks, irrespective of the ease of travel. A broader point that deserves future research is that while intuition suggests a correspondence in most places, connectivity may not always translate into human or pathogen movement.

Future directions for the work presented here include quantitative comparison and integration with other connectivity data, the integration of different pathogen weightings, and the extension to other regions of the World. Qualitative comparisons outlined above show some good correspondence with analyses of alternative sources of connectivity and disease data. A future step will be to compare these different connections and communities quantitatively to examine the weight of evidence for delineating areas of strong and weak connectivity. This could potentially follow similar studies looking at community structure on weighted networks, such as in the US based on commuting data \cite{nelson2016economic}, or UK and Belgium from mobile network data \cite{ratti2010redrawing,expert2011uncovering}. Here, \emph{P.falciparum} malaria was used to provide an example of the potential for weighting analyses by pathogen occurrence, prevalence, incidence or transmission suitability. Moreover, future work will examine the integration of alternative pathogen weightings. The maximum difference method was used here to pick out regions well connected to areas high \emph{P.falciparum} burden, but the potential exists to use different weighting methods depending on requirements, strategic needs, and the nature of the pathogen being studied.

Despite the rapid growth of air travel, shipping and rail in many parts of the world, roads continue to be the dominant route on which humans move on sub-national, national and regional scales. They form a powerful force in shaping the development of areas, facilitating trade and economic growth, but also bringing with them the exchange of pathogens. Results here show that their connectivity is not equal however, with strong clusters of high connectivity separated by bridge regions of low network density. These structures can have a significant impact on how pathogens spread, and by mapping them, a valuable evidence base to guide disease surveillance as well as control and elimination planning can be built.

%%%%%%%%%%%%%%%%%%%%%%%%%%%%%%%%%%%%%%%%%%%%%%%%%%%%%%%%%%%%%%%%%%%
%%%%%%%%%%%%%%%%%%%%%%%%%%%%%%%%%%%%%%%%%%%%%%%%%%%%%%%%%%%%%%%%%%%

\section*{Acknowledgments}
E.S. has been supported by funding from the Swiss National Science Foundation. 
A.J.T. is supported by funding from NIH/NIAID (U19AI089674), the Bill and Melinda Gates Foundation (OPP1106427, 1032350, OPP1134076), the Clinton Health Access Initiative, National Institutes of Health and a Wellcome Trust Sustaining Health Grant (106866/Z/15/Z).

\section*{Author contributions statement}

ES, MPV and AJT conceived and designed the analyses. ES and MPV designed the road network community mapping methods and undertook the analyses. All authors contributed to writing and reviewing the manuscript. 

\section*{Additional information}
\textbf{Competing financial interests} The authors declare no competing financial interests.

\bibliography{road_comm}

\begin{thebibliography}{10}
\expandafter\ifx\csname url\endcsname\relax
  \def\url#1{\texttt{#1}}\fi
\expandafter\ifx\csname urlprefix\endcsname\relax\def\urlprefix{URL }\fi
\expandafter\ifx\csname doiprefix\endcsname\relax\def\doiprefix{DOI }\fi
\providecommand{\bibinfo}[2]{#2}
\providecommand{\eprint}[2][]{\url{#2}}

\bibitem{tatem2006global}
\bibinfo{author}{Tatem, A.~J.}, \bibinfo{author}{Rogers, D.~J.} \&
  \bibinfo{author}{Hay, S.}
\newblock \bibinfo{journal}{\bibinfo{title}{Global transport networks and
  infectious disease spread}}.
\newblock {\emph{\JournalTitle{Advances in parasitology}}}
  \textbf{\bibinfo{volume}{62}}, \bibinfo{pages}{293--343}
  (\bibinfo{year}{2006}).

\bibitem{peiris2004severe}
\bibinfo{author}{Peiris, J.}, \bibinfo{author}{Guan, Y.} \&
  \bibinfo{author}{Yuen, K.}
\newblock \bibinfo{journal}{\bibinfo{title}{Severe acute respiratory
  syndrome}}.
\newblock {\emph{\JournalTitle{Nature medicine}}}
  \textbf{\bibinfo{volume}{10}}, \bibinfo{pages}{S88} (\bibinfo{year}{2004}).

\bibitem{webster2006h5n1}
\bibinfo{author}{Webster, R.~G.} \& \bibinfo{author}{Govorkova, E.~A.}
\newblock \bibinfo{journal}{\bibinfo{title}{H5n1 influenza—continuing
  evolution and spread}}.
\newblock {\emph{\JournalTitle{New England Journal of Medicine}}}
  \textbf{\bibinfo{volume}{355}}, \bibinfo{pages}{2174--2177}
  (\bibinfo{year}{2006}).

\bibitem{trifonov2009geographic}
\bibinfo{author}{Trifonov, V.}, \bibinfo{author}{Khiabanian, H.} \&
  \bibinfo{author}{Rabadan, R.}
\newblock \bibinfo{journal}{\bibinfo{title}{Geographic dependence,
  surveillance, and origins of the 2009 influenza a (h1n1) virus}}.
\newblock {\emph{\JournalTitle{New England journal of medicine}}}
  \textbf{\bibinfo{volume}{361}}, \bibinfo{pages}{115--119}
  (\bibinfo{year}{2009}).

\bibitem{MARAIS20163}
\bibinfo{author}{Marais, B.~J.}
\newblock \bibinfo{journal}{\bibinfo{title}{The global tuberculosis situation
  and the inexorable rise of drug-resistant disease}}.
\newblock {\emph{\JournalTitle{Advanced Drug Delivery Reviews}}}
  \textbf{\bibinfo{volume}{102}}, \bibinfo{pages}{3 -- 9}
  (\bibinfo{year}{2016}).

\bibitem{lynch2011transit}
\bibinfo{author}{Lynch, C.} \& \bibinfo{author}{Roper, C.}
\newblock \bibinfo{journal}{\bibinfo{title}{The transit phase of migration:
  circulation of malaria and its multidrug-resistant forms in africa}}.
\newblock {\emph{\JournalTitle{PLoS medicine}}} \textbf{\bibinfo{volume}{8}},
  \bibinfo{pages}{e1001040} (\bibinfo{year}{2011}).

\bibitem{hupalo2016population}
\bibinfo{author}{Hupalo, D.~N.} \emph{et~al.}
\newblock \bibinfo{journal}{\bibinfo{title}{Population genomics studies
  identify signatures of global dispersal and drug resistance in plasmodium
  vivax}}.
\newblock {\emph{\JournalTitle{Nature genetics}}}
  \textbf{\bibinfo{volume}{48}}, \bibinfo{pages}{953} (\bibinfo{year}{2016}).

\bibitem{tatem2012air}
\bibinfo{author}{Tatem, A.} \emph{et~al.}
\newblock \bibinfo{journal}{\bibinfo{title}{Air travel and vector-borne disease
  movement}}.
\newblock {\emph{\JournalTitle{Parasitology}}} \textbf{\bibinfo{volume}{139}},
  \bibinfo{pages}{1816--1830} (\bibinfo{year}{2012}).

\bibitem{tatem2014mapping}
\bibinfo{author}{Tatem, A.~J.}
\newblock \bibinfo{journal}{\bibinfo{title}{Mapping population and pathogen
  movements}}.
\newblock {\emph{\JournalTitle{International health}}}
  \textbf{\bibinfo{volume}{6}}, \bibinfo{pages}{5--11} (\bibinfo{year}{2014}).

\bibitem{lemey2014unifying}
\bibinfo{author}{Lemey, P.} \emph{et~al.}
\newblock \bibinfo{journal}{\bibinfo{title}{Unifying viral genetics and human
  transportation data to predict the global transmission dynamics of human
  influenza h3n2}}.
\newblock {\emph{\JournalTitle{PLoS pathogens}}} \textbf{\bibinfo{volume}{10}},
  \bibinfo{pages}{e1003932} (\bibinfo{year}{2014}).

\bibitem{Moustafa2017virome}
\bibinfo{author}{Moustafa, A.} \emph{et~al.}
\newblock \bibinfo{journal}{\bibinfo{title}{The blood dna virome in 8,000
  humans}}.
\newblock {\emph{\JournalTitle{PLOS Pathogens}}} \textbf{\bibinfo{volume}{13}},
  \bibinfo{pages}{1--20} (\bibinfo{year}{2017}).

\bibitem{tatem2012spatialHIV}
\bibinfo{author}{Tatem, A.~J.}, \bibinfo{author}{Hemelaar, J.},
  \bibinfo{author}{Gray, R.~R.} \& \bibinfo{author}{Salemi, M.}
\newblock \bibinfo{journal}{\bibinfo{title}{Spatial accessibility and the
  spread of hiv-1 subtypes and recombinants}}.
\newblock {\emph{\JournalTitle{Aids}}} \textbf{\bibinfo{volume}{26}},
  \bibinfo{pages}{2351--2360} (\bibinfo{year}{2012}).

\bibitem{Faria2014earlyHIV}
\bibinfo{author}{Faria, N.~R.} \emph{et~al.}
\newblock \bibinfo{journal}{\bibinfo{title}{The early spread and epidemic
  ignition of hiv-1 in human populations}}.
\newblock {\emph{\JournalTitle{Science}}} \textbf{\bibinfo{volume}{346}},
  \bibinfo{pages}{56--61} (\bibinfo{year}{2014}).

\bibitem{kraemer2017spread}
\bibinfo{author}{Kraemer, M.~U.} \emph{et~al.}
\newblock \bibinfo{journal}{\bibinfo{title}{Spread of yellow fever virus
  outbreak in angola and the democratic republic of the congo 2015--16: a
  modelling study}}.
\newblock {\emph{\JournalTitle{The Lancet Infectious Diseases}}}
  \textbf{\bibinfo{volume}{17}}, \bibinfo{pages}{330--338}
  (\bibinfo{year}{2017}).

\bibitem{dudas2017virus}
\bibinfo{author}{Dudas, G.} \emph{et~al.}
\newblock \bibinfo{journal}{\bibinfo{title}{Virus genomes reveal factors that
  spread and sustained the ebola epidemic}}.
\newblock {\emph{\JournalTitle{Nature}}} \textbf{\bibinfo{volume}{544}},
  \bibinfo{pages}{309--315} (\bibinfo{year}{2017}).

\bibitem{wesolowski2014commentary}
\bibinfo{author}{Wesolowski, A.} \emph{et~al.}
\newblock \bibinfo{journal}{\bibinfo{title}{Commentary: containing the ebola
  outbreak-the potential and challenge of mobile network data}}.
\newblock {\emph{\JournalTitle{PLoS currents}}} \textbf{\bibinfo{volume}{6}}
  (\bibinfo{year}{2014}).

\bibitem{scott2009world}
\bibinfo{author}{Scott, A.~J.}
\newblock \bibinfo{title}{World development report 2009: reshaping economic
  geography} (\bibinfo{year}{2009}).

\bibitem{Linard2012Africa}
\bibinfo{author}{Linard, C.}, \bibinfo{author}{Gilbert, M.},
  \bibinfo{author}{Snow, R.~W.}, \bibinfo{author}{Noor, A.~M.} \&
  \bibinfo{author}{Tatem, A.~J.}
\newblock \bibinfo{journal}{\bibinfo{title}{Population distribution, settlement
  patterns and accessibility across africa in 2010}}.
\newblock {\emph{\JournalTitle{PLOS ONE}}} \textbf{\bibinfo{volume}{7}},
  \bibinfo{pages}{1--8} (\bibinfo{year}{2012}).

\bibitem{Garrison1962}
\bibinfo{author}{{Garrison}, D., W.L.~{Marble}}.
\newblock \bibinfo{journal}{\bibinfo{title}{{The structure of transportation
  networks}}}.
\newblock {\emph{\JournalTitle{Techical report}}}  (\bibinfo{year}{1962}).

\bibitem{Haggett1969}
\bibinfo{author}{Haggett, P.} \& \bibinfo{author}{Chorley, R.~J.}
\newblock \emph{\bibinfo{title}{Network analysis in geography}},
  vol.~\bibinfo{volume}{67} (\bibinfo{publisher}{Edward Arnold London},
  \bibinfo{year}{1969}).

\bibitem{Barthelemy2011}
\bibinfo{author}{Barth{\'e}lemy, M.}
\newblock \bibinfo{journal}{\bibinfo{title}{Spatial networks}}.
\newblock {\emph{\JournalTitle{Physics Reports}}}
  \textbf{\bibinfo{volume}{499}}, \bibinfo{pages}{1--101}
  (\bibinfo{year}{2011}).

\bibitem{Strano2012}
\bibinfo{author}{Strano, E.}, \bibinfo{author}{Nicosia, V.},
  \bibinfo{author}{Latora, V.}, \bibinfo{author}{Porta, S.} \&
  \bibinfo{author}{Barth{\'e}lemy, M.}
\newblock \bibinfo{journal}{\bibinfo{title}{Elementary processes governing the
  evolution of road networks}}.
\newblock {\emph{\JournalTitle{Sci. Rep.}}} \textbf{\bibinfo{volume}{2}}
  (\bibinfo{year}{2012}).

\bibitem{strano_EPB_2013}
\bibinfo{author}{Strano, E.} \emph{et~al.}
\newblock \bibinfo{journal}{\bibinfo{title}{Urban street networks, a
  comparative analysis of ten european cities.}}
\newblock {\emph{\JournalTitle{Environment and Planning B: Planning and
  Design}}} \textbf{\bibinfo{volume}{40}}, \bibinfo{pages}{1071--1086}
  (\bibinfo{year}{2013}).

\bibitem{Strano170590}
\bibinfo{author}{Strano, E.} \emph{et~al.}
\newblock \bibinfo{journal}{\bibinfo{title}{The scaling structure of the global
  road network}}.
\newblock {\emph{\JournalTitle{Royal Society Open Science}}}
  \textbf{\bibinfo{volume}{4}} (\bibinfo{year}{2017}).

\bibitem{porta}
\bibinfo{author}{Porta, S.} \emph{et~al.}
\newblock \bibinfo{journal}{\bibinfo{title}{Street centrality and densities of
  retail and services in {B}ologna, {I}taly}}.
\newblock {\emph{\JournalTitle{Environ. Plann. B}}}
  \textbf{\bibinfo{volume}{36}}, \bibinfo{pages}{450--465}
  (\bibinfo{year}{2009}).

\bibitem{tatem2014integrating}
\bibinfo{author}{Tatem, A.~J.} \emph{et~al.}
\newblock \bibinfo{journal}{\bibinfo{title}{Integrating rapid risk mapping and
  mobile phone call record data for strategic malaria elimination planning}}.
\newblock {\emph{\JournalTitle{Malaria journal}}}
  \textbf{\bibinfo{volume}{13}}, \bibinfo{pages}{52} (\bibinfo{year}{2014}).

\bibitem{tatem2010international}
\bibinfo{author}{Tatem, A.~J.} \& \bibinfo{author}{Smith, D.~L.}
\newblock \bibinfo{journal}{\bibinfo{title}{International population movements
  and regional plasmodium falciparum malaria elimination strategies}}.
\newblock {\emph{\JournalTitle{Proceedings of the National Academy of
  Sciences}}} \textbf{\bibinfo{volume}{107}}, \bibinfo{pages}{12222--12227}
  (\bibinfo{year}{2010}).

\bibitem{wangdi2015chapter}
\bibinfo{author}{Wangdi, K.}, \bibinfo{author}{Gatton, M.~L.},
  \bibinfo{author}{Kelly, G.~C.} \& \bibinfo{author}{Clements, A.~C.}
\newblock \bibinfo{journal}{\bibinfo{title}{Cross-border malaria: A major
  obstacle for malaria elimination}}.
\newblock {\emph{\JournalTitle{Advances in parasitology}}}
  \textbf{\bibinfo{volume}{89}}, \bibinfo{pages}{79--107}
  (\bibinfo{year}{2015}).

\bibitem{OSM}
\bibinfo{author}{{Open Street Map}}.
\newblock \urlprefix\url{https://www.openstreetmap.org}.

\bibitem{GSM}
\bibinfo{author}{{Center for International Earth Science Information Network
  -CIESIN- Columbia University, Information Technology Outreach Services -ITOS-
  University of Georgia. Global Roads Open Access Data Set, Version 1
  (gROADSv1). Palisades, NY: NASA Socioeconomic Data and Applications Center
  (SEDAC)}} (\bibinfo{year}{2013}).
\newblock \urlprefix\url{http://dx.doi.org/10.7927/H4VD6WCT}.

\bibitem{GARMIN}
\bibinfo{author}{{GARMIN}}.
\newblock \urlprefix\url{https://developer.garmin.com/datasets/digital-atlas/}.

\bibitem{bhatt2015effect}
\bibinfo{author}{Bhatt, S.} \emph{et~al.}
\newblock \bibinfo{journal}{\bibinfo{title}{The effect of malaria control on
  plasmodium falciparum in africa between 2000 and 2015}}.
\newblock {\emph{\JournalTitle{Nature}}} \textbf{\bibinfo{volume}{526}},
  \bibinfo{pages}{207--211} (\bibinfo{year}{2015}).

\bibitem{Porta2006}
\bibinfo{author}{Porta, S.}, \bibinfo{author}{Crucitti, P.} \&
  \bibinfo{author}{Latora, V.}
\newblock \bibinfo{journal}{\bibinfo{title}{The network analysis of urban
  streets: a primal approach}}.
\newblock {\emph{\JournalTitle{Environment and Planning B: Planning and
  Design}}} \textbf{\bibinfo{volume}{33}}, \bibinfo{pages}{705}
  (\bibinfo{year}{2006}).

\bibitem{masucci2009random}
\bibinfo{author}{Masucci, A.~P.}, \bibinfo{author}{Smith, D.},
  \bibinfo{author}{Crooks, A.} \& \bibinfo{author}{Batty, M.}
\newblock \bibinfo{journal}{\bibinfo{title}{Random planar graphs and the london
  street network}}.
\newblock {\emph{\JournalTitle{The European Physical Journal B-Condensed Matter
  and Complex Systems}}} \textbf{\bibinfo{volume}{71}},
  \bibinfo{pages}{259--271} (\bibinfo{year}{2009}).

\bibitem{strano2013urban}
\bibinfo{author}{Strano, E.} \emph{et~al.}
\newblock \bibinfo{journal}{\bibinfo{title}{Urban street networks, a
  comparative analysis of ten european cities}}.
\newblock {\emph{\JournalTitle{Environment and Planning B: Planning and
  Design}}} \textbf{\bibinfo{volume}{40}}, \bibinfo{pages}{1071--1086}
  (\bibinfo{year}{2013}).

\bibitem{clauset2004finding}
\bibinfo{author}{Clauset, A.}, \bibinfo{author}{Newman, M.~E.} \&
  \bibinfo{author}{Moore, C.}
\newblock \bibinfo{journal}{\bibinfo{title}{Finding community structure in very
  large networks}}.
\newblock {\emph{\JournalTitle{Physical review E}}}
  \textbf{\bibinfo{volume}{70}}, \bibinfo{pages}{066111}
  (\bibinfo{year}{2004}).

\bibitem{Rosvall2005}
\bibinfo{author}{Rosvall, M.}, \bibinfo{author}{Trusina, A.},
  \bibinfo{author}{Minnhagen, P.} \& \bibinfo{author}{Sneppen, K.}
\newblock \bibinfo{journal}{\bibinfo{title}{Networks and cities: An information
  perspective}}.
\newblock {\emph{\JournalTitle{Physical Review Letters}}}
  \textbf{\bibinfo{volume}{94}}, \bibinfo{pages}{028701}
  (\bibinfo{year}{2005}).

\bibitem{Porta2006B}
\bibinfo{author}{Porta, S.}, \bibinfo{author}{Crucitti, P.} \&
  \bibinfo{author}{Latora, V.}
\newblock \bibinfo{journal}{\bibinfo{title}{The network analysis of urban
  streets: A dual approach}}.
\newblock {\emph{\JournalTitle{Physica A Statistical Mechanics and its
  Applications}}} \textbf{\bibinfo{volume}{369}}, \bibinfo{pages}{853--866}
  (\bibinfo{year}{2006}).

\bibitem{Newman06062006}
\bibinfo{author}{Newman, M. E.~J.}
\newblock \bibinfo{journal}{\bibinfo{title}{Modularity and community structure
  in networks}}.
\newblock {\emph{\JournalTitle{Proceedings of the National Academy of
  Sciences}}} \textbf{\bibinfo{volume}{103}}, \bibinfo{pages}{8577--8582}
  (\bibinfo{year}{2006}).

\bibitem{wesolowski2013use}
\bibinfo{author}{Wesolowski, A.} \emph{et~al.}
\newblock \bibinfo{journal}{\bibinfo{title}{The use of census migration data to
  approximate human movement patterns across temporal scales}}.
\newblock {\emph{\JournalTitle{PloS one}}} \textbf{\bibinfo{volume}{8}},
  \bibinfo{pages}{e52971} (\bibinfo{year}{2013}).

\bibitem{wesolowski2012quantifying}
\bibinfo{author}{Wesolowski, A.} \emph{et~al.}
\newblock \bibinfo{journal}{\bibinfo{title}{Quantifying the impact of human
  mobility on malaria}}.
\newblock {\emph{\JournalTitle{Science}}} \textbf{\bibinfo{volume}{338}},
  \bibinfo{pages}{267--270} (\bibinfo{year}{2012}).

\bibitem{tejedor2017travel}
\bibinfo{author}{Tejedor-Garavito, N.} \emph{et~al.}
\newblock \bibinfo{journal}{\bibinfo{title}{Travel patterns and demographic
  characteristics of malaria cases in swaziland, 2010--2014}}.
\newblock {\emph{\JournalTitle{Malaria Journal}}}
  \textbf{\bibinfo{volume}{16}}, \bibinfo{pages}{359} (\bibinfo{year}{2017}).

\bibitem{pindolia2012human}
\bibinfo{author}{Pindolia, D.~K.} \emph{et~al.}
\newblock \bibinfo{journal}{\bibinfo{title}{Human movement data for malaria
  control and elimination strategic planning}}.
\newblock {\emph{\JournalTitle{Malaria journal}}}
  \textbf{\bibinfo{volume}{11}}, \bibinfo{pages}{205} (\bibinfo{year}{2012}).

\bibitem{smith2017malaria}
\bibinfo{author}{Smith, J.~L.} \emph{et~al.}
\newblock \bibinfo{journal}{\bibinfo{title}{Malaria risk in young male
  travellers but local transmission persists: a case--control study in low
  transmission namibia}}.
\newblock {\emph{\JournalTitle{Malaria journal}}}
  \textbf{\bibinfo{volume}{16}}, \bibinfo{pages}{70} (\bibinfo{year}{2017}).

\bibitem{simon2013malaria}
\bibinfo{author}{Simon, C.} \emph{et~al.}
\newblock \bibinfo{journal}{\bibinfo{title}{Malaria control in botswana,
  2008--2012: the path towards elimination}}.
\newblock {\emph{\JournalTitle{Malaria journal}}}
  \textbf{\bibinfo{volume}{12}}, \bibinfo{pages}{458} (\bibinfo{year}{2013}).

\bibitem{raman2016reviewing}
\bibinfo{author}{Raman, J.} \emph{et~al.}
\newblock \bibinfo{journal}{\bibinfo{title}{Reviewing south africa’s malaria
  elimination strategy (2012--2018): progress, challenges and priorities}}.
\newblock {\emph{\JournalTitle{Malaria journal}}}
  \textbf{\bibinfo{volume}{15}}, \bibinfo{pages}{438} (\bibinfo{year}{2016}).

\bibitem{koita2013targeting}
\bibinfo{author}{Koita, K.} \emph{et~al.}
\newblock \bibinfo{journal}{\bibinfo{title}{Targeting imported malaria through
  social networks: a potential strategy for malaria elimination in swaziland}}.
\newblock {\emph{\JournalTitle{Malaria journal}}}
  \textbf{\bibinfo{volume}{12}}, \bibinfo{pages}{219} (\bibinfo{year}{2013}).

\bibitem{lynch2015association}
\bibinfo{author}{Lynch, C.~A.} \emph{et~al.}
\newblock \bibinfo{journal}{\bibinfo{title}{Association between recent internal
  travel and malaria in ugandan highland and highland fringe areas}}.
\newblock {\emph{\JournalTitle{Tropical medicine \& international health}}}
  \textbf{\bibinfo{volume}{20}}, \bibinfo{pages}{773--780}
  (\bibinfo{year}{2015}).

\bibitem{pearce2009multiple}
\bibinfo{author}{Pearce, R.~J.} \emph{et~al.}
\newblock \bibinfo{journal}{\bibinfo{title}{Multiple origins and regional
  dispersal of resistant dhps in african plasmodium falciparum malaria}}.
\newblock {\emph{\JournalTitle{PLoS medicine}}} \textbf{\bibinfo{volume}{6}},
  \bibinfo{pages}{e1000055} (\bibinfo{year}{2009}).

\bibitem{guimera2005worldwide}
\bibinfo{author}{Guimera, R.}, \bibinfo{author}{Mossa, S.},
  \bibinfo{author}{Turtschi, A.} \& \bibinfo{author}{Amaral, L.~N.}
\newblock \bibinfo{journal}{\bibinfo{title}{The worldwide air transportation
  network: Anomalous centrality, community structure, and cities' global
  roles}}.
\newblock {\emph{\JournalTitle{Proceedings of the National Academy of
  Sciences}}} \textbf{\bibinfo{volume}{102}}, \bibinfo{pages}{7794--7799}
  (\bibinfo{year}{2005}).

\bibitem{kaluza2010complex}
\bibinfo{author}{Kaluza, P.}, \bibinfo{author}{K{\"o}lzsch, A.},
  \bibinfo{author}{Gastner, M.~T.} \& \bibinfo{author}{Blasius, B.}
\newblock \bibinfo{journal}{\bibinfo{title}{The complex network of global cargo
  ship movements}}.
\newblock {\emph{\JournalTitle{Journal of the Royal Society Interface}}}
  \textbf{\bibinfo{volume}{7}}, \bibinfo{pages}{1093--1103}
  (\bibinfo{year}{2010}).

\bibitem{APMEN}
\bibinfo{author}{{APMEN, Asian Pacific Malaria Elimination Network}}.
\newblock \urlprefix\url{http://apmen.org}.

\bibitem{SADC}
\bibinfo{author}{{SADC, Southern African Development Community }}.
\newblock
  \urlprefix\url{https://tis.sadc.int/english/sarn/elimination-eight-e8}.

\bibitem{sorichetta2016mapping}
\bibinfo{author}{Sorichetta, A.} \emph{et~al.}
\newblock \bibinfo{journal}{\bibinfo{title}{Mapping internal connectivity
  through human migration in malaria endemic countries}}.
\newblock {\emph{\JournalTitle{Scientific data}}} \textbf{\bibinfo{volume}{3}},
  \bibinfo{pages}{160066} (\bibinfo{year}{2016}).

\bibitem{ruktanonchai2016census}
\bibinfo{author}{Ruktanonchai, N.~W.} \emph{et~al.}
\newblock \bibinfo{journal}{\bibinfo{title}{Census-derived migration data as a
  tool for informing malaria elimination policy}}.
\newblock {\emph{\JournalTitle{Malaria journal}}}
  \textbf{\bibinfo{volume}{15}}, \bibinfo{pages}{273} (\bibinfo{year}{2016}).

\bibitem{marshall2016key}
\bibinfo{author}{Marshall, J.~M.} \emph{et~al.}
\newblock \bibinfo{journal}{\bibinfo{title}{Key traveller groups of relevance
  to spatial malaria transmission: a survey of movement patterns in four
  sub-saharan african countries}}.
\newblock {\emph{\JournalTitle{Malaria journal}}}
  \textbf{\bibinfo{volume}{15}}, \bibinfo{pages}{200} (\bibinfo{year}{2016}).

\bibitem{bradley2015infection}
\bibinfo{author}{Bradley, J.} \emph{et~al.}
\newblock \bibinfo{journal}{\bibinfo{title}{Infection importation: a key
  challenge to malaria elimination on bioko island, equatorial guinea}}.
\newblock {\emph{\JournalTitle{Malaria journal}}}
  \textbf{\bibinfo{volume}{14}}, \bibinfo{pages}{46} (\bibinfo{year}{2015}).

\bibitem{nelson2016economic}
\bibinfo{author}{Nelson, G.~D.} \& \bibinfo{author}{Rae, A.}
\newblock \bibinfo{journal}{\bibinfo{title}{An economic geography of the united
  states: from commutes to megaregions}}.
\newblock {\emph{\JournalTitle{PloS one}}} \textbf{\bibinfo{volume}{11}},
  \bibinfo{pages}{e0166083} (\bibinfo{year}{2016}).

\bibitem{ratti2010redrawing}
\bibinfo{author}{Ratti, C.} \emph{et~al.}
\newblock \bibinfo{journal}{\bibinfo{title}{Redrawing the map of great britain
  from a network of human interactions}}.
\newblock {\emph{\JournalTitle{PloS one}}} \textbf{\bibinfo{volume}{5}},
  \bibinfo{pages}{e14248} (\bibinfo{year}{2010}).

\bibitem{expert2011uncovering}
\bibinfo{author}{Expert, P.}, \bibinfo{author}{Evans, T.~S.},
  \bibinfo{author}{Blondel, V.~D.} \& \bibinfo{author}{Lambiotte, R.}
\newblock \bibinfo{journal}{\bibinfo{title}{Uncovering space-independent
  communities in spatial networks}}.
\newblock {\emph{\JournalTitle{Proceedings of the National Academy of
  Sciences}}} \textbf{\bibinfo{volume}{108}}, \bibinfo{pages}{7663--7668}
  (\bibinfo{year}{2011}).

\end{thebibliography}

%%%%%%%%%%%%%%%%%%%%%%%%%%%%%%%%%%%%%%%%%%%%%%%%%%%%%%%%%%%%%%%%%%%
%% FIGURES %%%%%%%%%%%%%%%%%%%%%%%%%%%%%%%%%%%%%%%%%%%%%%%%%%%%%%%

\end{document}